\documentclass{elsart3-1}

\usepackage{graphicx}
\usepackage{epsfig}

\usepackage{amssymb}

\usepackage[english,francais]{babel}

\newtheorem{e-proposition}[theorem]{Proposition}

\newtheorem{e-definition}[theorem]{Definition\rm}

\renewcommand{\theequation}{\arabic{equation}}
\def\ba{\begin{eqnarray}}
\def\ea{\end{eqnarray}}
\def\be{\begin{equation}}  
\def\ee{\end{equation}}
\def\bea{\begin{eqnarray}}
\def\eea{\end{eqnarray}}
\def\beqn{\begin{eqnarray}}
\def\eeqn{\end{eqnarray}}
\def\beq{\begin{equation}}  
\def\eeq{\end{equation}}
\def\abs#1{\left|#1\right|} 
\newcommand\sss{\scriptscriptstyle}
\def \lsim{\mathrel{\vcenter
     {\hbox{$<$}\nointerlineskip\hbox{$\sim$}}}}

\def\og{\leavevmode\raise.3ex\hbox{$\scriptscriptstyle\langle\!\langle$~}}
\def\fg{\leavevmode\raise.3ex\hbox{~$\!\scriptscriptstyle\,\rangle\!\rangle$}}

\begin{document}
\centerline{Higgs boson properties \ldots /
Propri\'et\'es du boson de Higgs\ldots}
\begin{frontmatter}


\selectlanguage{english}
\title{Higgs Boson Properties in the Standard Model \\ 
and its Supersymmetric Extensions}


\selectlanguage{english}
\author[authorlabel1]{John~Ellis},
\ead{John.Ellis@cern.ch}
\author[authorlabel2]{Giovanni~Ridolfi}, 
\ead{Giovanni.Ridolfi@ge.infn.it}
\author[authorlabel3]{Fabio~Zwirner}
\ead{Fabio.Zwirner@pd.infn.it}

\address[authorlabel1]{Theory Division, Physics Department, CERN,
CH-1211 Geneva 23, Switzerland}
\address[authorlabel2]{Dipartimento di Fisica, Universit\`a di Genova and INFN,
Sezione di Genova, Via Dodecaneso 33, I-16146 Genova, Italy}
\address[authorlabel3]{Dipartimento di Fisica, Universit\`a di Padova and INFN,
Sezione di Padova, Via Marzolo 8, I-35131 Padova, Italy}

\begin{abstract}
We review the realization of the Brout-Englert-Higgs mechanism in the
electroweak theory and describe the experimental and theoretical
constraints on the mass of the single Higgs boson expected in the
minimal Standard Model. We also discuss the couplings of this Higgs
boson and its possible decay modes as functions of its unknown
mass. We then review the structure of the Higgs sector in the minimal
supersymmetric extension of the Standard Model (MSSM), noting the
importance of loop corrections to the masses of its five physical
Higgs bosons. Finally, we discuss some non-minimal models.\\
{\it To cite this article: J.~Ellis, G.~Ridolfi, F.~Zwirner,
C. R. Physique X (2007).}

\vskip 0.5\baselineskip

\selectlanguage{francais}
\noindent{\bf R\'esum\'e}
\vskip 0.5\baselineskip
\noindent
{\bf Propri\'et\'es du boson de Higgs dans le Mod\`ele Standard
et ses extensions supersym\'etriques}\\
\noindent
Nous examinons le m\'ecanisme Brout-Englert-Higgs et d\'ecrivons les
limites exp\'erimentales et th\'eoriques de la masse de l'unique boson
de Higgs attendu dans le Mod\`ele Standard minimal.  Nous discutons
\'egalement les couplages de ce boson de Higgs et ses modes de
d\'esint\'egration en fonction de sa masse inconnue.  Nous examinons
ensuite la structure du secteur de Higgs dans l'extension
supersym\'etrique minimale du Mod\`ele Standard (MSSM), en soulignant
l'importance des corrections induites par les boucles aux masses de
ses cinq Higgs bosons physiques.  Enfin, nous examinons quelques
mod\`eles non-minimaux.\\ {\it Pour citer cet article~: J.~Ellis,
G.~Ridolfi, F.~Zwirner, C. R. Physique X (2007).}
\bigskip

\begin{center}
CERN-PH-TH/2007-012
\qquad
DFPD-07-TH-03
\qquad
GEF-TH-10/2007
\end{center}

\vskip 0.5\baselineskip
\noindent{\small{\it Mots-cl\'es~:} Boson de Higgs; Brisure de la sym\'etrie electrofaible;
Supersym\'etrie}}
\end{abstract}
\end{frontmatter}

\selectlanguage{english}
\section{Introduction}
\label{intro}

The key properties of any elementary particle are its mass, its spin and 
its couplings to other particles. Owing to its very specific  theoretical 
r\^ole~\cite{precursors,BEH,Higgs}, the single Higgs boson~\cite{Higgs}
of the 
Standard Model of strong and electroweak interactions \cite{SM} has  
some very characteristic properties. In particular, its spin is zero,  
unlike any other elementary particle ever observed, and its couplings  
to other particles are proportional to their masses. 
The mass  dependence of its couplings is related to the special r\^ole 
that this scalar field has in generating the masses of the other elementary  
particles.

On the other hand, the mass of the Higgs boson is largely
unconstrained in the Standard Model, although there are
upper bounds derived from unitarity \cite{unitarity}. More recently,
significant lower limits have been provided by direct experimental
searches at LEP \cite{lephiggs}, and precision electroweak data seem
to prefer a specific portion of the mass range allowed by these upper and
lower limits \cite{EWWG}. Moreover, if one requires the Standard Model
to remain valid up to high energies, there are upper and
lower limits \cite{thbounds} on the possible mass of the Higgs
boson. Nevertheless, within the Standard Model the mass of the Higgs
boson remains an unknown quantity, though it may be more constrained  in
certain extensions such as supersymmetry. As a result, knowing the
couplings of the the Standard Model Higgs boson enables one to
predict its decay properties only as functions of its unknown mass.

Later we use the standard field-theoretical point of view to discuss the properties 
of the Standard Model Higgs boson. However, they can be
understood qualitatively using very simple arguments, without appealing to the
details of its field-theoretical formulation. In order to be compatible with 
unitarity, the cross sections for particle scattering are bounded, and 
cannot grow without limit. Moreover, a field theory is capable of making 
many predictions, over a wide range of energy scales and in terms of a 
finite number of input parameters, only if it is renormalizable, i.e., if the 
divergences encountered in calculating quantum loop diagrams can be 
absorbed in the definitions of the input parameters. The requirement of
renormalizability imposes further restrictions on scattering cross sections.

The scattering cross sections in the Standard Model are
bounded suitably at high energies only if the Higgs boson
is included, and this requirement determines its properties uniquely~\cite{cancel}. 
For example, if
one considers the annihilation of a fermion-antifermion pair into a
pair of $W^\pm$ bosons, it is well known that the high-energy
behaviour of the cross section is largely tamed by cancellations
between direct-channel photon- and $Z$-exchange diagrams with the
crossed-channel neutrino-exchange diagram, an effect measured at LEP
\cite{threevectors}. However, there is in principle a residual
divergence proportional to the product of the fermion and $W$ masses,
which occurs only in the direct-channel spin-0 partial
wave\footnote{This was numerically insignificant at LEP, because of
the very small electron mass.  However, it would be a very large
effect in ${\bar t}t \to W^+ W^-$ annihilation.}. The only way to
remove this residual divergence is to include the direct-channel
exchange of a spin-0 particle whose couplings to fermions and $W$
bosons are determined by their masses~\cite{cancel}.  Similar divergences would
occur in $W^+ W^- \to W^+ W^-$ scattering, a process that will be
observable at the LHC. Again, the only way to tame this divergence
within the Standard Model is to include the exchange of a scalar Higgs
boson with a coupling to the $W$ boson that is proportional to its
mass~\cite{cancel}.

The Higgs sector of the Standard Model that is described in the next
Section of this paper is the simplest field-theoretical realization of
these cancellations, and it is known to be part of a renormalizable
theory \cite{rensm}. After formulating the theory, we then discuss the
couplings and decay modes of the single Higgs boson of the Standard
Model~\cite{EGN,hunter}, as well as theoretical, phenomenological,
experimental and cosmological bounds on its possible mass. The
following Section extends this discussion to supersymmetric extensions
of the Standard Model at the Fermi scale of weak interactions, as
motivated by requiring~\cite{hiesusy} the gauge hierarchy to be
natural~\cite{hierarchy}. We discuss supersymmetric extensions of the
Standard Model, mainly the minimal one (MSSM), but also some
non-minimal versions.

There are five physical Higgs bosons in the MSSM, three of them
neutral and two charged.  All of them have spin 0, and the couplings
of the neutral ones obey sum rules that reflect how they share the
mass-generating task of the single Higgs boson of the Standard Model,
and the subtle interplay of gauge symmetry and
supersymmetry~\cite{hunter}. The masses and couplings of these Higgs
bosons are calculable in terms of underlying model parameters, and
receive important quantum corrections~\cite{erz,oyy,hh}.
Nevertheless, at least in the MSSM, the mass of the lightest Higgs
boson is tightly circumscribed to be less than about 130~GeV, and its
couplings are generally predictable. As a result, in quite a large
part of the MSSM parameter space, its properties are expected to be
quite similar to those of a light Standard Model Higgs boson. If a
light Higgs boson is found at the LHC, it will be a challenge to
distinguish between the Standard Model and its minimal supersymmetric
extension, unless one or more heavier Higgs bosons or some
supersymmetric particles are also found. On the other hand, if no
light Higgs boson is found at the LHC, it would be difficult to
maintain faith in supersymmetry at the weak scale as a solution to the
gauge-hierarchy problem.

\section{Standard Model}
\label{SM}

\subsection{General properties}

In this section, we review the Brout-Englert-Higgs mechanism~\cite{BEH,Higgs}
in the minimal version of the Standard Model (SM from now on). An
$SU(2)$-doublet scalar field $\phi$ is introduced, which is allowed to
transform inhomogeneously under the action of the SU(2) factor of the
SM gauge group:
\beq
\phi=\left(\begin{array}{c} \phi_1 \\ \phi_2 \end{array}\right) \, ,
\qquad
\phi\to e^{ig\xi^a\frac{\sigma^a}{2}}\left(\phi+\frac{v}{\sqrt{2}}\right)
-\frac{v}{\sqrt{2}} \, ,
\qquad
v=\left(\begin{array}{c}v_1\\v_2\end{array}\right) \, ,
\eeq
where $g$ is the $SU(2)$ coupling constant, $\xi^a$ are real
transformation parameters, $\sigma^a$ ($a=1,2,3$) are the Pauli
matrices, and $v_1$ and $v_2$ are scalar constants.  Such an
inhomogeneous transformation rule is possible only for a scalar field
$\phi$: otherwise, the values of $v_1,v_2$ would depend on the choice
of reference frame. The value of the hypercharge of $\phi$ is fixed by
the requirement that it transform homogeneously under ordinary
electric charge transformations that correspond to the subgroup
U(1)$_{\rm em}$.  This amounts to requiring
\beq
e^{ie\alpha Q}\,\frac{1}{\sqrt{2}}
\left(\begin{array}{c} v_1 \\ v_2 \end{array}\right)
-\frac{1}{\sqrt{2}}
\left(\begin{array}{c} v_1 \\ v_2 \end{array}\right)=0 \, ,
\eeq
or equivalently
\beq
\left(\begin{array}{cc} Q_1 & 0\\0& Q_2\end{array}\right)
\left(\begin{array}{c} v_1 \\ v_2 \end{array}\right)
=\left(\begin{array}{cc} 1/2+Y/2& 0\\0& -1/2+Y/2\end{array}\right)
\left(\begin{array}{c} v_1 \\ v_2 \end{array}\right)
=\left(\begin{array}{c} 0 \\ 0 \end{array}\right) \, ,
\label{noemb}
\eeq
where $Q_{1,2}$ are the electric charges of $\phi_{1,2}$ in units of the proton 
charge $e$, and we have used $Q=T_3+Y/2$. The system in eq.~(\ref{noemb}) 
has non-vanishing solutions only for $Y=\pm 1$. Without loss of generality, we 
shall adopt the first choice, $Y=+1$, which gives $Q_1=1,Q_2=0$. The solution 
of eq.~(\ref{noemb}) is in this case $v_1=0, v_2\equiv v$, and we can also 
assume without loss of generality that $v$ is real and positive.

The covariant derivative
\be
D_\mu\left(\phi+\frac{v}{\sqrt{2}}\right)=\left(\partial_\mu
-ig\frac{\sigma^a}{2}W_\mu^a-ig'\frac{1}{2}B_\mu\right)
\left(\phi+\frac{v}{\sqrt{2}}\right)
\ee
transforms as an ordinary doublet of SU(2), and mass terms for the gauge vector 
boson fields arise in the term $\abs{D_\mu(\phi+v/\sqrt{2})}^2$, with
$m_W^2=g^2v^2/4$, $m_Z^2=(g^2+{g'}^2)v^2/4$, whereas the photon is massless.

The form of the scalar potential is uniquely determined by 
renormalizability and gauge invariance, together with the requirement
that $\phi=0$ correspond to a minimum:
\beq
V(\phi)=\lambda\left[\left(\phi+\frac{v}{\sqrt{2}}\right)^\dag
\left(\phi+\frac{v}{\sqrt{2}}\right)
-\frac{v^2}{2}\right]^2.
\eeq
Three of the four scalar degrees of freedom in the doublet $\phi$ are
unphysical, because of the gauge symmetry; they do not appear in
$S$-matrix elements as asymptotic states. Their presence as
intermediate states is needed in order to cancel unphysical
singularities in the gauge boson propagators, but they may be removed
from the spectrum by choosing a unitary gauge.  The fourth degree of
freedom is instead a physical one; we shall denote the corresponding
real scalar field by $H$. It corresponds to a massive, spinless
neutral particle, the Higgs boson~\cite{Higgs},
with squared mass $m_H^2=2\lambda v^2$.

The two constants $v$ and $\lambda$ are fundamental parameters of the
theory, whose values must be extracted from experiment. The value
$v\simeq 250$~GeV can be obtained from the measured value of the Fermi
constant, $G_F/\sqrt{2}=g^2/(8m_W^2)=1/( 2 v^2)$,
whereas the value of $\lambda$ (and hence of $m_H$) is still unknown.

Fermion mass terms arise from Yukawa interactions with the Higgs
doublet.  We do not reproduce here the details of this mechanism, but
we do recall that, whereas the Yukawa couplings of the SM fermions to
the Higgs doublet are in general non-diagonal in the space of
electroweak doublets of fermions, as are the corresponding couplings
of the $W^\pm$ gauge bosons, the couplings of the neutral $Z$ boson
and the single physical Higgs boson of the SM are diagonal in flavour
space at the tree level, and flavour-changing phenomena induced by
neutral currents are strongly suppressed in loop
amplitudes~\cite{gim}, as required by experiment.  Violation of CP
invariance can also be implemented with three or more fermion families
\cite{km}, using just one Higgs doublet~\cite{EGNCP}.


The couplings and self-couplings of the Higgs boson are contained in the
following interaction terms of the SM Lagrangian:
\bea
{\mathcal L}_{\rm Higgs}&=&
\left(m_{\sss W}^2\,{W^\mu}^+W_\mu^-
+\frac{1}{2}\,m_{\sss Z}^2\, Z^\mu Z_\mu\right)
\left(\frac{H^2}{v^2}+\frac{2H}{v}\right)
-\frac{H}{v}\sum_f m^f\,\bar \psi^f \psi^f
-\lambda v H^3-\frac{1}{4}\lambda H^4 \, ,
\eea
where the $\psi^f$ are Dirac fields for massive fermions, and the $m^f$ are 
the corresponding masses. We see that the theory contains couplings of
the Higgs boson to the $W,Z$ bosons that are proportional to $m^2_V/v\sim g
m_V$ ($V=W,Z$), and Yukawa couplings to fermions that are proportional to
$m^f/v\sim g m^f/m_V$, as argued earlier on the basis of unitarity and
renormalizability. 

\begin{figure}[ht]
\begin{center}
\includegraphics[width=.90\textwidth]{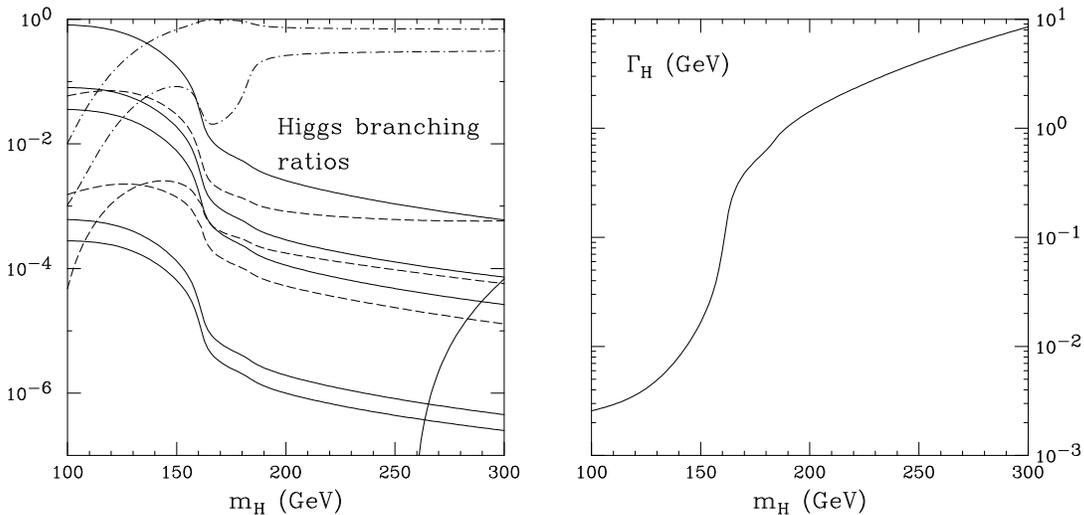}
\caption{\em Left: branching ratios of the different Higgs decay modes.
Solid: $H\to f\bar f$ ($f=t,b,c,\tau,s,\mu$); dot-dashed: $WW,ZZ$; 
dashed: $gg,\gamma\gamma, Z\gamma$.
Right: total width of the Higgs boson as a function of its mass.
This figure was obtained with the program HDECAY~\protect\cite{hdecay}.}
\label{fig:higgsproperties}
\end{center}
\end{figure}

It follows that, for $m_H<2m_W$, the Higgs
boson decays mainly into pairs of heavy fermions ($b\bar b$ or
$\tau^+\tau^-$), while for higher values of $m_H$, the $W^+W^-$ and
$ZZ$ channels open and become dominant~\cite{unitarity}. At even larger values of
$m_H$, the $t\bar t$ mode is also available, though it never becomes 
as important as the decays into pairs of gauge bosons. The rare decay into a
pair of photons~\cite{EGN} is also potentially important for detection at the LHC, 
where the coupling to gluon pairs~\cite{wilczek} may be important for Higgs production~\cite{GGMN}.
Both these couplings are induced at the one-loop level by loops of
heavy electromagnetically or colour charged particles. Below the $W^+W^-$ threshold, the $gg$
branching ratio is comparable to those for the $\tau^+ \tau^-$ and $
c \bar c$ channels. The branching ratios for the various decay modes
are displayed in the left panel of fig.~\ref{fig:higgsproperties}, as
functions of the Higgs mass.
The total width of the Higgs boson is shown in the right panel of
fig.~\ref{fig:higgsproperties} as a function of the Higgs mass. It is
below 1 GeV for $m_H\lsim 200$~GeV, but becomes comparable to $m_H$
for $m_H\sim 1$~TeV.

\subsection{Theoretical bounds}

In principle, the self-interaction coupling constant $\lambda$ of the Higgs
scalar can take any (real and positive) value; thus, the mass of the physical 
Higgs boson is not fixed by the theory. However, as anticipated in the 
Introduction, some information about the value of $m_H$ can be obtained 
by means of theoretical considerations.

Perturbative unitarity \cite{unitarity} provides an upper bound on the Higgs 
boson mass, much in the same way as in the case of the $W$ boson. The 
process to be considered is the elastic scattering of weak vector bosons 
with longitudinal polarizations; the unitarity bound on the 
leading-order amplitude is respected, including the contribution of Higgs 
boson intermediate states, provided the value of the Higgs mass
does not exceed $800-1000$~GeV~\cite{unitarity}. A similar bound can be obtained 
by considering Higgs decays, whose rates for $WW$ and $ZZ$ final states
grow as $m_H^3$ for large $m_H$, and by requiring that the total width
$\Gamma_H$ is smaller than $m_H$. Note that a value of $m_H$ close 
to the unitarity bound would imply a large value of the coupling $\lambda$:
non-perturbative phenomena would in this case become important, and
the whole mechanism should be reconsidered in this light, since a 
perturbative approach becomes unreliable.

\begin{figure}[ht]
\begin{center}
\includegraphics[width=.65\textwidth]{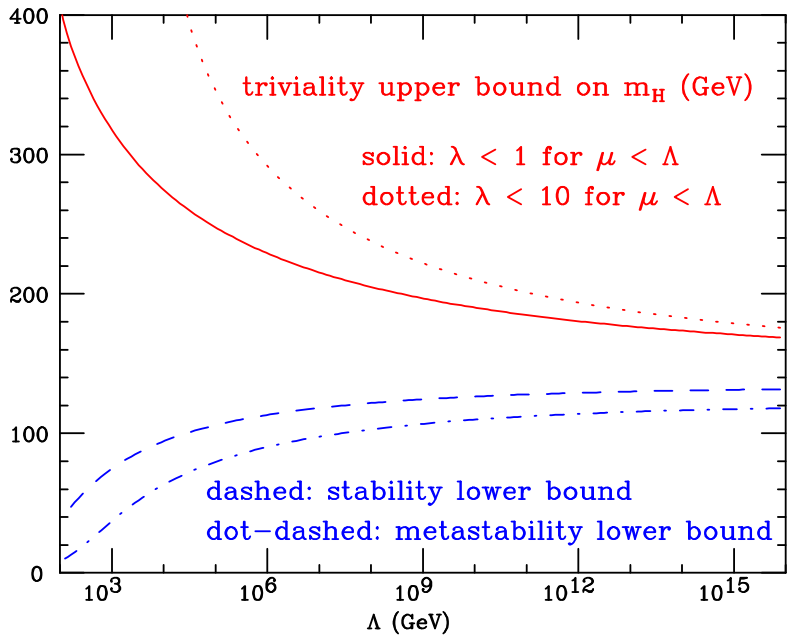}
\caption{\em Theoretical bounds on the Higgs boson mass
as a function of the energy scale~\protect\cite{irs}.}
\label{fig:bounds}
\end{center}
\end{figure}

A second class of theoretical bounds arises from the study of the
behaviour of the quartic coupling $\lambda$ under the effect of
renormalization~\cite{thbounds}. For sufficiently small values of the
Higgs boson mass, $m_H^2 \simeq 2 \lambda(v) v^2$, the solution of the
renormalization-group equation for $\lambda$ is a decreasing function
of the renormalization scale $\mu$ for $\mu$ of the order of the weak
scale, and eventually becomes negative at a value $\Lambda$ which
depends on the initial condition for $\lambda$ at the weak scale,
$\mu=v$: the smaller $\lambda(v)$, the smaller $\Lambda$. If
$\lambda(\mu)<0$, the effective potential is not bounded from below,
and the ground state becomes unstable.  Therefore, the theory is only
consistent at energy scales $\mu\lsim\Lambda$.  Correspondingly, $m_H$
has a $\Lambda$-dependent lower bound, which is shown in
fig.~\ref{fig:bounds}, taken from a recent analysis~\cite{irs}.
A slightly less restrictive lower bound is obtained by allowing metastability
of the ground state, with the constraint that its lifetime be larger than
the age of the universe. This metastability bound is also shown in
fig.~\ref{fig:bounds}.

The argument can be reversed: some new degrees of
freedom would have to show their effects at energy scales of the order of
$\Lambda$, in order to restore the stability of the ground state.
The value of $\Lambda$ is therefore of great importance, since it
is related to the energy scale at which we should expect non-standard
phenomena to take place.

For larger values of $m_H^2\simeq 2\lambda(v) v^2$, 
the running coupling $\lambda(\mu)$ increases at larger scales,
and has a simple (Landau) pole in $\log\mu$, similarly to what happens 
in electrodynamics. This behaviour is easily seen in the perturbative 
solution for the running coupling, and is confirmed by non-perturbative 
(lattice) computations (see, e.g., \cite{lattice} and references therein). The 
theory cannot then be consistent up to arbitrarily large energies. 
Thus, one may view the SM as an effective theory,
valid up to some energy scale $\Lambda$, such that $\lambda(\mu)$ remains
small for $\mu\lsim \Lambda$. Since the larger the value of $\lambda(v)$, the smaller
the value of $\Lambda$, a $\Lambda$-dependent upper bound on $m_H$ is obtained.
This bound (usually referred to as the triviality bound) is also
shown in fig.~\ref{fig:bounds}, for two different choices of the maximum
value for $\lambda$.

We see from fig.~\ref{fig:bounds} that if the SM
is assumed to be valid up to very large energy scales,
of the order of the extrapolated unification scale for the gauge
coupling constants~\cite{gqw}, a relatively small allowed range
for the Higgs boson mass is obtained.

\subsection{Direct searches and constraints from precision measurements}

Most of our present experimental knowledge about the SM Higgs boson
comes from the study of $e^+e^-$ collisions performed at LEP and the
SLC between 1988 and 2000. As concerns the Higgs boson, the results of
this enormous amount of experimental work can be summarized as
follows. No direct evidence for the existence of the SM Higgs has been
produced. This allows one to set a lower limit on the Higgs mass of
114.4~GeV, mainly based of the non-observation of Higgs bosons in
association with a $Z^0$~\cite{EGN,vh,hunter}, followed by the decay of the Higgs into a
heavy fermion-antifermion pair \cite{lephiggs}.

On the other hand, the value of the Higgs boson mass affects the SM
predictions for the observables measured in experiments at LEP, the
SLC, the Tevatron collider and elsewhere, through radiative
corrections. The impact of the Higgs boson is relatively small;
furthermore, the dependence of the precision observables on the Higgs
mass is only logarithmic at one loop (as opposed, for example, to the
dependence on the top quark mass, which is quadratic)~\cite{Veltman}.

\begin{figure}
\begin{center}
\includegraphics[width=.7\textwidth]{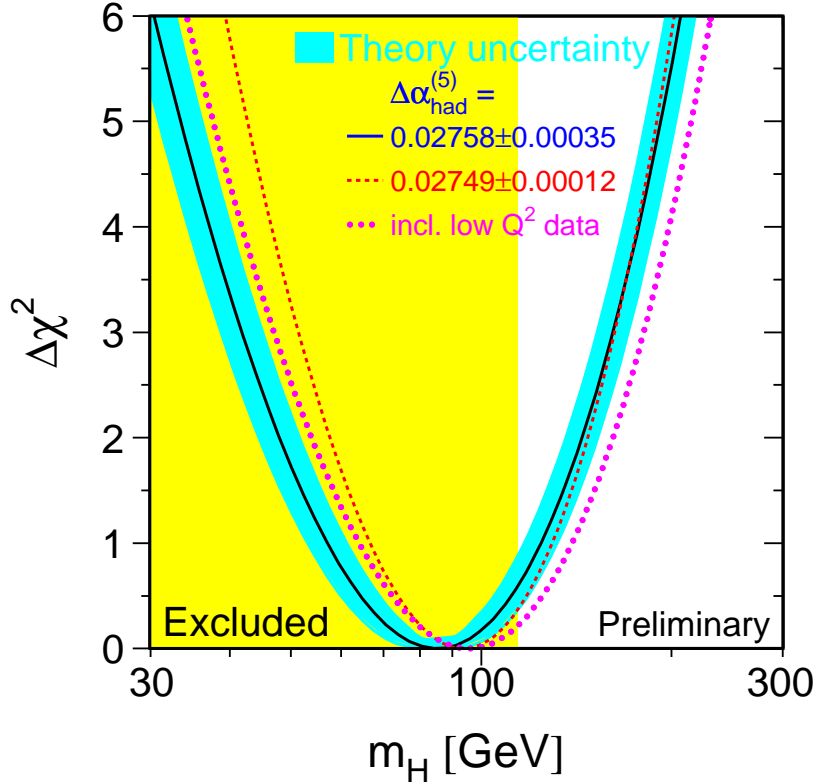}
\caption{{\it The value of $\Delta\chi^2=\chi^2-\chi^2_{\rm min}$
as a function of $m_H$, used as a fit parameter, for a global fit
to precision observables in the SM~\protect\cite{EWWG}.}}
\label{fig:chisquare}
\end{center}
\end{figure}

However, thanks to the great accuracies achieved by the electroweak
experiments (some of the observables have been measured with an
accuracy of the order of 0.1\%), it is possible to perform a global
fit~\cite{EWWG} using the SM Higgs mass as a free parameter, thus
obtaining a preferred value for $m_H$ and the corresponding
uncertainty~\footnote{It should be recalled that a similar procedure
was followed with the top quark mass as a free parameter, before the
direct observation of the top quark at the Tevatron. The fitted value
of $m_t$ was very close to the observed value of about $175$~GeV.}.
The result of such a global fit is displayed in
Fig.~\ref{fig:chisquare},
where the value of $\Delta \chi^2=\chi^2-\chi^2_{\rm min}$ is shown as
a function of $m_H$. The solid black curve and the dotted red curve
correspond to different evaluations of the renormalization of the
electromagnetic coupling due to light quarks, and the dashed purple
line includes more low-$Q^2$ data. The blue band represents the
uncertainty due to higher-order effects. At the $\Delta\chi^2=1$
level, the global electroweak fit yields~\cite{EWWG} $\log_{10} m_H
({\rm GeV}) =1.93^{+0.16}_{-0.17}$, or
\begin{equation}
m_H \; = \; 85^{+39}_{-28}~{\rm GeV},
\end{equation}
which is consistent with the bounds discussed in the previous
section. It should be noted, however, that the value of $m_H$
preferred by the precision observables is slightly below the present
exclusion limit.

\begin{figure}[ht]
\begin{center}
\includegraphics[width=.45\textwidth]{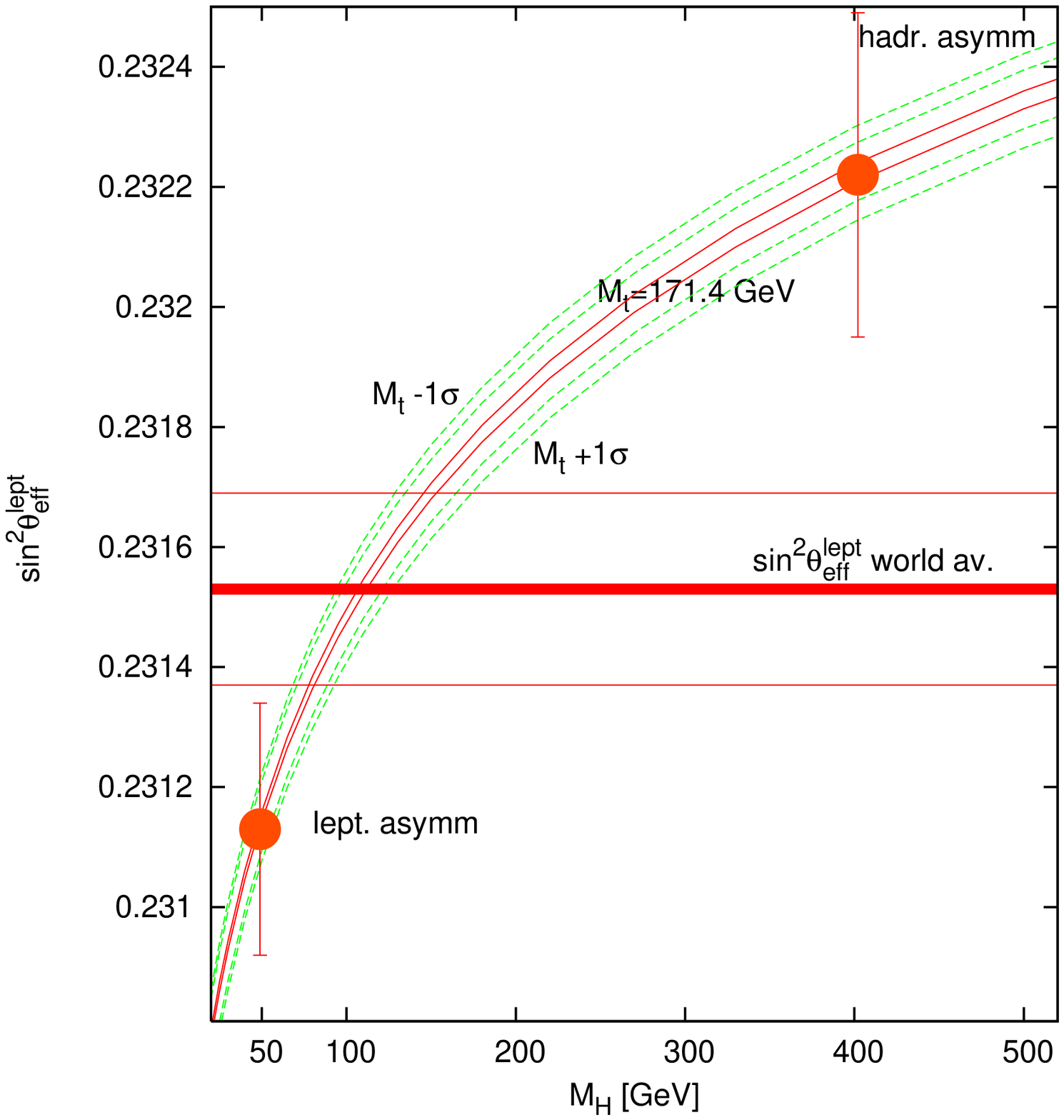}
\includegraphics[width=.45\textwidth]{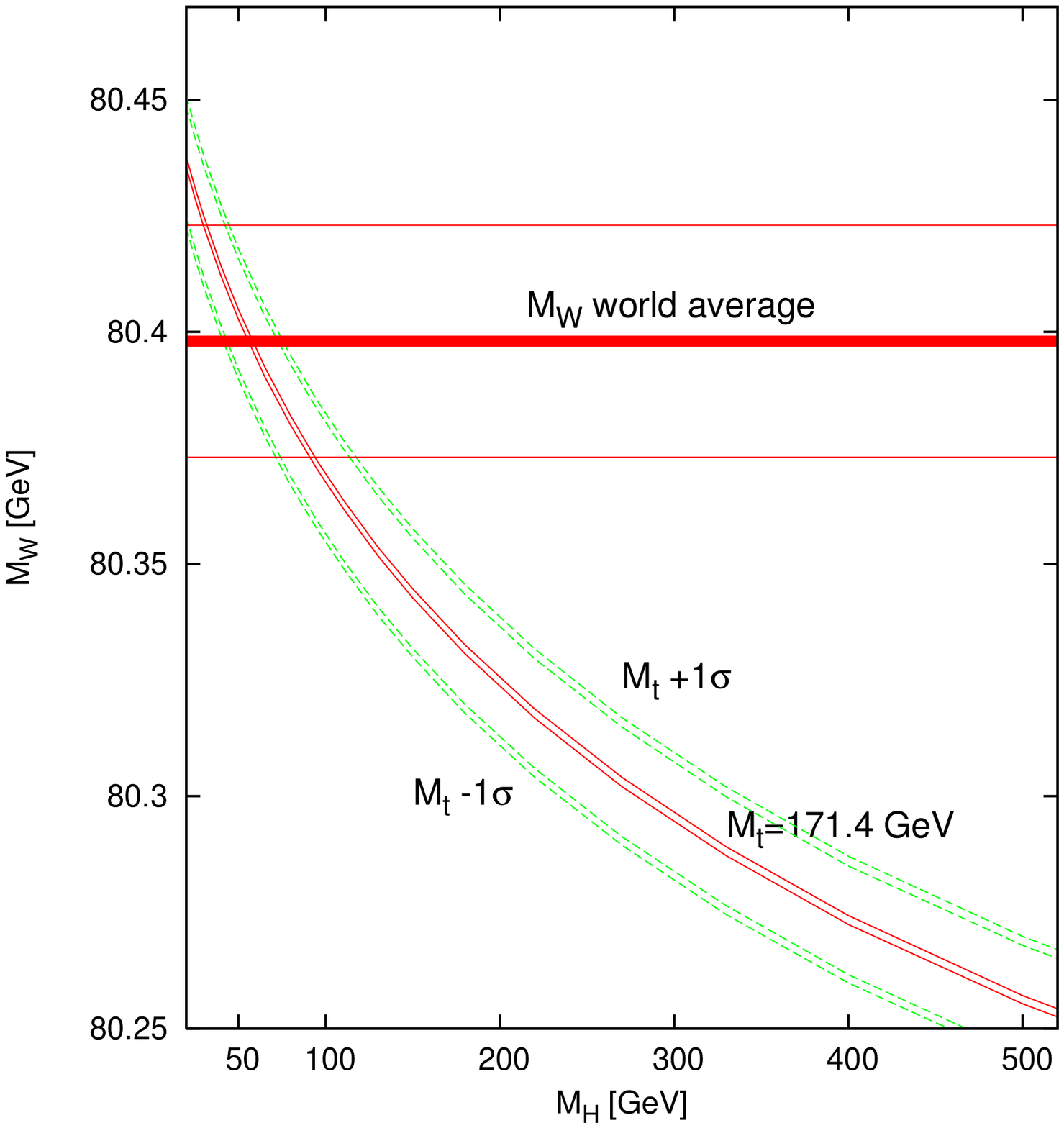}
\caption{\em Impact of different observables on
the fit to $m_H$ (updated from~\protect\cite{gambino}).}
\label{fig:paolo1}
\end{center}
\end{figure}

The fit is reasonably good, but not excellent: the value of $\chi^2$
per degree of freedom is 17.8/13, corresponding to a probablity of $17 \%$.  
This poorness of fit is mainly due to a marginal
discrepancy between forward-backward asymmetries for hadronic final
states, which favour $m_H$ around $400$ GeV, and
leptonic asymmetries, which prefer a lower value~\cite{Chanowitz}.  This effect
is illustrated in the left panel of Fig.~\ref{fig:paolo1}, where the
computed value of the weak mixing angle is shown as a function of
$m_H$, with the top mass fixed at its central value or at its lower
and upper bounds.
The mass of the $W$ boson also indicates a relatively small value of the Higgs
mass, as one can see in the right panel of fig.~\ref{fig:paolo1},
where we show the computed value of the $W$ boson mass as a function
of $m_H$. The measured value of $m_W$ is also shown, with its present
error band. Indeed, within the SM, this single observable would prefer a value 
of $m_H$ below the direct LEP exclusion limit.

Higgs searches at the Tevatron are not sensitive yet to the small
cross sections valid for the SM Higgs \cite{tevhiggs}, but the situation
may change with the accumulation of sufficient integrated luminosity.
On the other
hand, the LHC should be sensitive to the SM Higgs in its entire
possible mass range \cite{lhchiggs}. We note here only that the most
important SM Higgs production mechanisms are expected to be
gluon-gluon~\cite{GGMN} and $WW/ZZ$ fusion \cite{VV}. Production in
association with a W/Z boson \cite{vh,GNY} or a top-antitop pair
\cite{httb} is suppressed but may still play an important role for
detection.

\section{Supersymmetric Models}
\label{susy}

\subsection{Motivations}

There are deep and general reasons to envisage supersymmetric
extensions of the SM (for a recent review with references to the
original literature, see, e.g.,~\cite{susyrev}). The most general symmetries 
of local relativistic quantum field theories include supersymmetry, 
and the link it provides between bosons and fermions is the
only type of symmetry that has not yet been discovered
in the observed fundamental interactions.  Supersymmetry also appears
to be an essential feature of string theory, our best
candidate for a unified quantum theory of all interactions. Finally,
supersymmetry provides, in its linear realization, a rationale for the
existence of elementary scalars, since they are combined with chiral fermions
in the same supersymmetric representations. 

The failure to find evidence for supersymmetry in experiments performed
so far implies that it must be broken. Arguments that are not fundamental, but nevertheless very
plausible, hint at an effective supersymmetry-breaking scale
very close to the present experimental bounds, with the superpartners
of the SM particles having masses around the Fermi scale of weak
interactions, $G_F^{-1/2} \simeq 300 \, {\rm GeV}$. The first reason is the
so-called hierarchy problem~\cite{hierarchy}: in the SM, there is no symmetry
protecting the Higgs mass parameter from quantum corrections quadratic
in the ultraviolet cut-off scale of the theory, $\Lambda$.
The scale invariance of the classical equations of motions, recovered in the limit of
vanishing Higgs mass parameter, is expected to be broken not only by
SM quantum effects, but also by the new physics at the cut-off scale.
Therefore, extrapolating the SM to scales $\Lambda$ much higher
than the Fermi scale requires an increasingly unnatural fine-tuning of
parameters. Supersymmetric field theories possess very special quantum
properties, including the absence of most quadratic and some logarithmic
quantum corrections~\cite{nrt}. Consequently, weak-scale supersymmetry
would suppress the sensitivity to the ultraviolet scale, and hence
might explain the stability of the hierarchy~\cite{hiesusy} between
the Fermi scale and the Planck scale, $M_P = 1/\sqrt{8 \pi G_N}$. This
feature enables supersymmetry also to provide a framework for
understanding the dynamical origin of the hierarchy. Additionally, the
renormalization-group evolution of the gauge couplings~\cite{gqw}
is changed in theories
with weak-scale
supersymmetry~\cite{susyrge}, which facilitates the grand unification of the gauge
couplings measured in low-energy experiments~\cite{GUTs}.
Supersymmetry at the electroweak scale also predicts a relatively light Higgs
boson, as favoured by the precision electroweak
data~\cite{EWWG}. Moreover, in many supersymmetric models, the
Lightest Supersymmetric Particle (LSP), typically a weakly-interacting
neutral spin-1/2 fermion called a neutralino, is stable and a suitable
candidate for the dark matter inferred from astrophysical and
cosmological data~\cite{EHNOS}.

In the absence of a satisfactory dynamical understanding of
supersymmetry breaking at a fundamental level (which cannot avoid
addressing the vacuum energy problem, with its unexplained huge
hierarchy between the vacuum energy scale, $10^{-3}$-$10^{-4}$~eV, and
the Planck scale), most phenomenological aspects of supersymmetry are
usually discussed within the minimal supersymmetric extension of the
Standard Model (MSSM) \cite{fayet,dg}. This is the most economical and
predictive supersymmetric extension of the SM at the weak scale, but
is certainly not the only possibility. In the following we describe in
some detail the properties of the MSSM Higgs sector, and conclude with
some comments on non-minimal models.

\subsection{The MSSM Higgs sector}

A general property of any (renormalizable) supersymmetric extension of
the SM is the presence of at least two Higgs doublets~\cite{fayet},
which leads to an extended Higgs sector. This is because, in the linear
realization of $N=1$ supersymmetry, spin-0 particles come in chiral
supermultiplets, whose physical degrees of freedom are a complex
spin-0 boson and a two-component spin-1/2 fermion. Trying to build a
realistic supersymmetric extension of the SM with just one Higgs
chiral supermultiplet $H$ would immediately lead to a triple
problem. First, there would be a massless charged fermion in the 
spectrum, since charged particles have Dirac masses,
involving an even number of two-component spinors. Furthermore, 
since in supersymmetric models Yukawa couplings are
determined by the superpotential, an analytic function of the chiral
superfields, it would be impossible to write down superpotential
couplings giving masses to all charged quarks and
leptons. Finally, the spin-1/2 component of $H$ would contribute
to the chiral anomaly, whose cancellation is required for the quantum
consistency of the theory.  As a result, any (renormalizable) 
supersymmetric extension of the SM must contain at
least two Higgs doublets with opposite weak hypercharge, denoted by
\begin{equation}
H_1 \equiv \left( \begin{array}{c} 
H_1^0 \\ H_1^- \end{array} \right) \sim 
(1,2,-1) \, ,
\qquad \qquad
H_2 \equiv \left( \begin{array}{c} 
H_2^+ \\ H_2^0 \end{array} \right) \sim 
(1,2,+1) \, ,
\label{doublets}
\end{equation}
where the SU(3) $\times$ SU(2) $\times$ U(1) quantum numbers are given
in brackets in an obvious notation. In contrast with a generic
two-Higgs-doublet model, supersymmetry ensures that, in the MSSM, the
$H_1^0$ has tree-level Yukawa couplings only with charge-1/3 quarks
and charged leptons, and the $H_2^0$ only with charge-2/3 quarks, enforcing
the natural suppression of flavour-changing neutral currents~\cite{gwp} in the
limit of exact supersymmetry.

For the moment, we concentrate on the MSSM, which is defined by the 
following properties: 1) the minimal gauge group, SU(3) $\times$ SU(2) $\times$ 
U(1); 2) the minimal particle content, three generations of quarks and leptons 
and two Higgs doublets plus their superpartners; 3) an exact discrete
R-parity, which guarantees baryon- and lepton-number conservation in
renormalizable interactions since $R=+1$ for SM particles and Higgs bosons
whereas $R=-1$ for their superpartners; 4) supersymmetry breaking parametrized
by explicit but soft breaking terms, comprising gaugino and scalar mass terms and
trilinear scalar couplings, without additional CP-violating phases
beyond those of the Yukawa couplings.

The tree-level potential of the MSSM depends as follows on the two Higgs doublets
in eq.~(\ref{doublets}):
\begin{equation}
V_0 = m_1^2 |H_1|^2 +  m_2^2 |H_2|^2 +  m_3^2 (H_1 H_2 + h.c.) + 
\frac{g^2}{8} (H_2^\dagger \sigma^a H_2 +  H_1^\dagger \sigma^a H_1)^2 
+  \frac{g^{\prime \, 2}}{8} (|H_2|^2 -  |H_1|^2)^2 \, ,
\label{vzero}
\end{equation}
where $m_1^2,m_2^2,m_3^2$ are mass parameters, $g$ and $g'$ are the
gauge coupling constants of SU(2) and U(1), respectively.
Eq.~(\ref{vzero})
displays a crucial difference between the SM and MSSM potentials: in
the SM the quartic coupling is an arbitrary parameter $\lambda$,
proportional to the SM Higgs mass whereas in the MSSM, because of
supersymmetry and in spite of the presence of a second Higgs doublet,
the quartic scalar couplings in $V_0$ are related to the
electroweak gauge couplings.

For suitable values of the mass parameters, $V_0$ has a minimum for
$\langle H_1^0 \rangle = v_1 \ne 0$, $\langle H_2^0 \rangle = v_2 \ne
0$, where it is not restrictive to take $v_1$ and $v_2$ real and
positive. The combination $v^2 = v_1^2 + v_2^2$ controls the Fermi
scale and the weak boson masses, and is fixed by experimental
data. Fermion masses are proportional to $v_2$ for charge-2/3 quarks,
and to $v_1$ for charge-1/3 quarks and charged leptons.  Neutrino masses
can be accounted for by modifications of the MSSM that do not affect
the present discussion. Of the eight real degrees of freedom of
eq.~(\ref{doublets}), three are the would-be Goldstone bosons that
provide the longitudinal components of the massive gauge bosons. The
physical spectrum contains then three neutral states, two of which are CP-even
$(h,H)$ and one CP-odd $(A)$, and two charged states $(H^\pm)$. 

\subsection{Tree-Level Masses and Couplings}

At the
classical level, all masses and couplings of the MSSM Higgs sector
depend only on measured SM parameters and two more independent
parameters: the latter are usually taken to be $\tan \beta \equiv
v_2/v_1$ and one mass parameter, for example $m_A$. Although $m_A$ is
essentially unconstrained, apart from naturalness arguments suggesting
that it should not be much larger than the Fermi scale, the range of
$\tan \beta$ favoured by model calculations is $1 < \tan \beta <
m_t/m_b$, where $m_t$ and $m_b$ are the running top and bottom masses
evaluated near the Fermi scale. The remaining tree-level mass
eigenvalues are then given by:
\begin{equation}
m_{H^\pm}^2 = m_W^2 + m_A^2 \, ,
\qquad
m_{h,H}^2 = \frac{1}{2} \left[ m_A^2 + m_Z^2 \mp
 \sqrt{(m_A^2+m_Z^2)^2 - 4 m_A^2 m_Z^2 \cos^2 2 \beta} \right]
\, , 
\label{clmass}
 \end{equation}
leading to stringent inequalities such as $m_W, m_A <
m_{H^{\pm}}$, $m_h < m_Z |\cos 2 \beta | < m_Z < m_H$, and $m_h < m_A <
m_H$, holding at the classical level. 

Similarly, all tree-level Higgs boson couplings can be computed
in terms of the mixing angle $\alpha$ that is required to diagonalize
the mass matrix of the neutral CP-even Higgs bosons, and is given by
\begin{equation}
\cos 2 \alpha = - \cos 2 \beta \; { { m_A^2  - m_Z^2 }\over {
 m_H^2 - m_h^2  }} \ , \qquad  - {\pi \over 2} < \alpha\  {\leq}\  0 \, .
\end{equation}
For example, the couplings of the three neutral MSSM Higgs bosons to 
vector-boson and fermion pairs are easily obtained from the SM Higgs
couplings, by multiplying the latter by the $\alpha$- and $\beta$-dependent 
factors summarized in Table~1.
\begin{table}
\begin{center}
\begin{tabular}{|c|c|c|c|}
\hline
& & & \\
&
$
d \overline{d} \, , \, l^+l^-
$
& $u\overline{u}$
& \ $W^+W^-,ZZ$ \  \\
& & & \\
\hline
& & & \\
\ $h$ \
& \ $- \sin \alpha / \cos \beta $ \
& \ $  \cos \alpha / \sin \beta $ \
& \ $  \sin \, (\beta -\alpha)  $ \ \\
& & & \\
\hline
& & & \\
\ $H$ \
& \ $\cos \alpha / \cos \beta$ \
& \ $\sin \alpha / \sin \beta$ \
& \ $\cos \, (\beta -\alpha)  $ \
\\
& & & \\
\hline
& & & \\
\ $A$ \
& $-i\gamma_5 \tan \beta $
& $-i\gamma_5 \cot \beta $
& $0$
\\
& & & \\
\hline
\end{tabular}
\caption{Correction factors, relative to their SM values,
for the couplings of the MSSM neutral
Higgs bosons to fermion and vector boson pairs, expressed in terms of the angle $\beta$
that describes the ratio of MSSM Higgs vacuum expectation values and the angle $\alpha$
that describes mixing between the CP-even neutral Higgs bosons.}
\end{center}
\end{table}
The remaining tree-level Higgs boson couplings in the MSSM can be
found, for example, in~\cite{hunter}. An interesting situation in
parameter space is the so-called decoupling limit: when $m_A^2 \gg
m_Z^2$, the lightest neutral Higgs boson $h$ behaves 
much as the SM Higgs boson, with $\alpha \sim
(\beta - \pi/2)$, whereas the $(H,A,H^\pm)$ are much heavier and nearly 
degenerate, forming an isospin doublet that decouples at sufficiently 
low energy.

\subsection{Radiative Corrections}

An important consequence of the structure of $V_0$ in
eq.~(\ref{vzero}) is the existence of at least one neutral CP-even
Higgs boson with mass smaller than $m_Z$ ($h$) or very close to it
($H$), and significantly coupled with the $W$ and $Z$ bosons. This
initially raised the hope that a crucial test of the MSSM Higgs
sector could be performed at LEP. However, as 
reviewed below, the inclusion of radiative corrections to the MSSM
Higgs sector~\cite{erz,oyy,hh} drastically changes the picture by
increasing $m_h$, and it was not possible to complete the test of the MSSM at LEP,
because its maximum centre-of-mass
energy was $\sqrt{s} = 209 \, {\rm
GeV}$, not much above $2 \, m_Z$.

Radiative corrections to the MSSM Higgs sector are dominated by loop
diagrams involving virtual top quarks ($t$) and their superpartners,
the spin-0 stop squarks, denoted by ($\tilde{t}_1,\tilde{t}_2$) in the
basis of definite mass. The leading one-loop effects can easily be
computed from the effective potential, and used for a first estimate
of the one-loop corrected mass eigenvalues and of the mixing angle
$\alpha$ in the neutral CP-even sector. Neglecting mixing in the stop
mass matrix, working in the decoupling limit $m_A^2 \gg m_Z^2$, and 
assuming $m_{\tilde{t}_1}^2, m_{\tilde{t}_2}^2 \gg m_t^2$ and $\tan \beta 
\ll m_t/m_b$, the most important correction to the lightest Higgs boson
mass takes the very simple form~\cite{erz,oyy,hh}:
\begin{equation}
\label{leading}
\Delta m_h^2  \simeq \frac{3 \, g^2 \, m_t^4}{16 \pi^2 \, m_W^2}
\log \frac{m_{\tilde{t}_1}^2 m_{\tilde{t}_2}^2}{m_t^4} \, .
\end{equation}
An alternative way to understand the size of this radiative correction
is to consider an effective theory in which all the heavy MSSM
particles, including the stop squarks, the extra Higgs bosons and the
top quark, have been integrated out: in this case, the quartic Higgs coupling in the
low-energy effective theory gets large positive contributions from
one-loop stop and top diagrams, which translate into a corresponding
increase in the Higgs boson mass. Mixing in the stop sector can lead
to further, large positive contributions to $m_h^2$. At one-loop
order, and again for simplicity in the decoupling limit, one finds, in
addition to (\ref{leading}):
\begin{equation}
(\Delta m_h^2)_{mix}  \simeq \frac{3 \, g^2 \, m_t^2 \, s^2_{2 \theta} \,
( m_{\tilde{t}_1}^2 - m_{\tilde{t}_2}^2)}{32 \pi^2 \, m_W^2} \left[
\log \frac{m_{\tilde{t}_1}^2}{m_{\tilde{t}_2}^2} +
\frac{ s^2_{2 \theta} \, ( m_{\tilde{t}_1}^2 - m_{\tilde{t}_2}^2)}{4 \, m_t^2}
\left( 1 - \frac{1}{2} \frac{ m_{\tilde{t}_1}^2 + m_{\tilde{t}_2}^2}{ m_{\tilde{t}_1}^2 
- m_{\tilde{t}_2}^2} \log \frac{m_{\tilde{t}_1}^2}{m_{\tilde{t}_2}^2}  \right) 
\right] \, ,
\end{equation}
where $\theta$ is the mixing angle in the stop mass matrix and
$s^2_{2\theta} \equiv [\sin (2 \theta)]^2$. 

\begin{figure}[ht]
\begin{center}
\includegraphics[width=.65\textwidth]{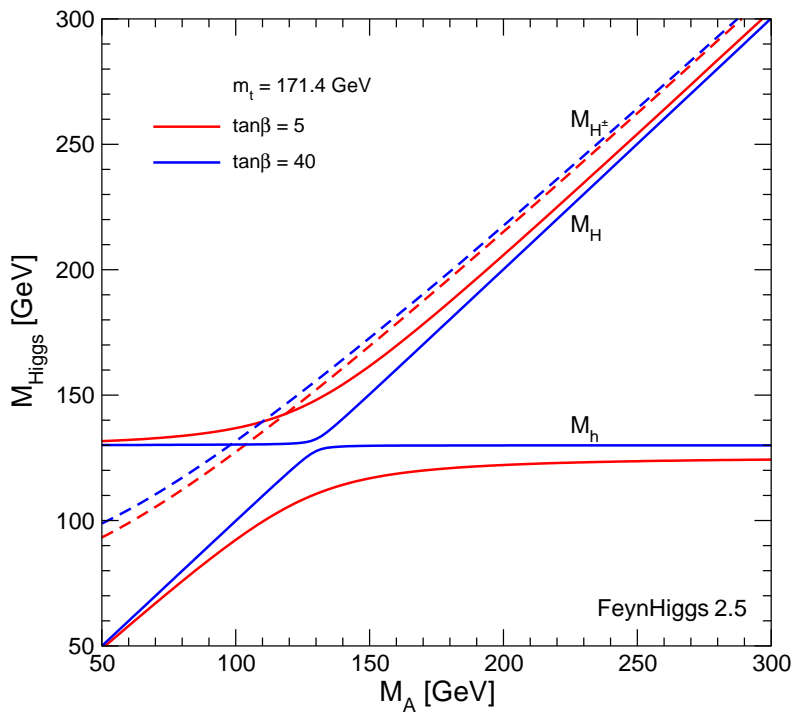}
\caption{\em Calculations of the lightest Higgs boson $h$, the heavier
CP-even neutral Higgs boson $H$ and the charged Higgs bosons $H^\pm$
as functions of the mass of the CP-odd Higgs boson $A$, as calculated
in the MSSM with a squark mass of 1~TeV, a gluino mass of 800~GeV, a
wino mass of 200~GeV, $A = \mu/\tan \beta + 2000$~GeV and $\mu =
200$~GeV, assuming $m_t = 171.4$~GeV, provided by S.~Heinemeyer
using the first of refs.~\protect\cite{Feynhiggs}.}
\label{fig:Feynhiggs}
\end{center}
\end{figure}

Over the years, extensive efforts have been devoted to progressive
refinements of these and other radiative corrections to the MSSM Higgs
sector, with special emphasis on the prediction for $m_h$. These
include resummation of large logarithms using one- and two-loop
renormalization group equations and calculation of all the most
important two-loop contributions. The status of the two-loop
calculations is presented in~\cite{susyrev}: the most important ones
are $O(\alpha_s \alpha_t)$ corrections, which tend to decrease the
upper bound on $m_h$, and $O(\alpha_t^2)$ corrections, which can
partially compensate the previous effect.  The most important
two-loop calculations are implemented in~\cite{Feynhiggs}, and we note
that the leading three-loop corrections have recently been shown to be
small \cite{3loop}.  Phenomenological analyses within the MSSM point
at a typical upper bound $m_h^{max} \sim 130 \ {\rm GeV}$ as seen in
Fig.~\ref{fig:Feynhiggs}: calculations performed within specific
models for supersymmetry breaking tend to produce lower values of
$m_h^{max}$, whereas stretching the model parameters beyond their
natural values, and taking the uncertainty in the top quark mass into
full account, can produce slightly higher values of $m_h^{max}$.
Radiative corrections also affect the MSSM Higgs boson couplings. For 
moderate values of $\tan \beta$, the leading corrections can be included by 
using tree-level formulae with running couplings, evaluated at an appropriate
scale, and the loop-corrected values of the mixing angles 
$\alpha$ and $\beta$. For large values of $\tan \beta$. however, there can
also be important threshold corrections to the Higgs couplings to the 
bottom/sbottom sector, with strong dependences on the model parameters. 
 
\subsection{Signals and phenomenology}

The experimental signatures of the MSSM Higgs bosons depend on their
branching ratios and on their production cross sections at high-energy
colliders. A systematic discussion can be very complicated, due to the
large number of parameters involved, already at the classical level
and even more so after the inclusion of radiative corrections. As for
the branching ratios, we content ourselves by mentioning some general
features that may distinguish the lightest MSSM Higgs boson from the
SM one, when the remaining MSSM Higgs bosons are significantly heavier
than the $Z$ but not yet completely decoupled. Being typically lighter
than 130 GeV, $h$ mostly decays into fermion pairs. Since $\tan \beta > 1$
in general, the branching
ratios for $h$ decays into $b \overline{b}$ and $\tau^+ \tau^-$ tend to be enhanced with
respect to a SM Higgs boson of the same mass.
Decays of $h$ into $AA$, or into final states with an even number of
supersymmetric particles, may be important in the regions of parameter
space where they are kinematically allowed.

In $e^+ e^-$ collisions, the main production mechanisms for the MSSM
neutral CP-even Higgs bosons are $e^+ e^- \rightarrow h Z \ (HZ)$ and
$e^+ e^- \rightarrow h A \ (HA)$. Vector-boson fusion can be important
for centre-of-mass energies much larger than the Higgs mass. This mechanism
did not play an important role at LEP, but it could play a role at a
future high-energy linear collider. At LEP, the relevant final states
were those generated by the intermediate states $hZ$ and $hA$, which
play a complementary role, since their cross-sections are proportional
to $\sin^2 (\beta - \alpha)$ and $\cos^2 (\beta - \alpha)$,
respectively.  Extensive searches have been carried out by the four
LEP collaborations and are summarized in the reports of the Higgs
Working group (for the most recent one at the time of this writing,
see~\cite{lephwg}). Values of
$m_h$ and $m_A$ below 92-93 GeV are typically excluded at the 95$\%$
c.l. over most of the MSSM parameter space.  The limit on $m_h$
gradually approaches that in the SM in the decoupling limit.

The present Tevatron searches for the lightest MSSM Higgs boson
\cite{tevhiggs} are marginally sensitive to the region of very large
$\tan \beta \sim m_t /m_b$, when the Yukawa couplings to $b$ quarks
and $\tau$ leptons are strongly enhanced with respect to the SM, but 
there are also strong constraints from rare B decays. Studies \cite{lhchiggs} 
indicate that the LHC will be sensitive to most of the MSSM parameter 
space, but distinguishing the light MSSM Higgs from the SM Higgs will 
not always be possible, in the absence of signals for other MSSM particles.

\subsection{Non-minimal models}

While CP cannot be spontaneously broken by the tree-level MSSM
Higgs potential, there is still the possibility of radiatively induced 
CP-violating effects in the MSSM Higgs sector~\cite{cpv}, coming from 
explicit CP-violating phases in other sectors of the MSSM. This leads 
to further complications in the discussion of the MSSM Higgs searches,
since all three neutral Higgs bosons can mix.

In the constrained MSSM with universal soft supersymmetry-breaking 
parameters, there are just two additional explicit CP-violating phases, 
namely the phases of the (common) gaugino mass $m_{1/2}$ and soft 
supersymmetry-breaking trilinear coupling $A$ relative to the Higgs 
mixing parameter $\mu$. Without loss of generality, one may assume 
that $\mu$ is real, and term the CP-violating phases $\theta_{1/2}$ and 
$\theta_A$. Their first effect on the Higgs sector of the MSSM is the 
generation of off-diagonal mass terms mixing the CP-even Higgs 
bosons $h, H$ with the CP-odd boson $A$~\cite{CPHiggs}, leading 
to a general $3\times 3$ Higgs-boson mass matrix.  Each of the
off-diagonal CP-violating scalar-pseudoscalar mixing contributions to the 
neutral MSSM mass-squared matrix contains terms scaling qualitatively as
\begin{equation}
\label{MSP}
M^2_{SP} \ \sim \ \frac{m^4_t}{v^2} \, 
\frac{\sin \theta_A |\mu| |A_t|}{32\pi^2\, M_{SUSY}^2}
\  \left( 6,\ \frac{|A_t|^2}{M_{SUSY}^2} \,, \ 
\frac{|\mu|^2}{\tan\beta \, M_{SUSY}^2} \,,\ 
\frac{2 \cos \theta_A |\mu| |A_t|}{M_{SUSY}^2} \,\right)\, ,
\end{equation}
where $A_t$ is the effective trilinear coupling associated with the top 
squarks, and $M_{SUSY}$ is an average stop mass. For $\theta_A \sim 
\pi/2$ and $|\mu|, |A_t| > M_{SUSY}$, $M^2_{SP}$ could be of order 
$M^2_Z$, though one should take into account the stringent constraints 
coming from electric dipole moments~\cite{OPRS}. The gaugino mass phase $\theta_{1/2}$ 
also has an influence at the two-loop level, via the gluino mass. In 
addition to inducing three-way mixing between the neutral Higgs bosons, the phases $\theta_A$ 
and $\theta_{1/2}$ also introduce CP-violating effects in the Higgs couplings. For codes to 
evaluate such effects, see~\cite{CPsuperH} and the first paper in Ref.~\cite{Feynhiggs}.

There is a sum rule requiring the sum of the squares of the three couplings of
neutral Higgs bosons to vector boson pairs to equal the square of the single 
Higgs-vector-vector coupling in the Standard Model, whereas this squared 
coupling is shared between just the two CP-even Higgs bosons in the CP-conserving 
case displayed in Table~1. Observable manifestations of CP violation in the MSSM 
Higgs sector have been considered at LEP, the LHC and other accelerators~\cite{CPHiggs}.
It is possible that the lightest neutral Higgs boson may be lighter than the LEP lower limit
obtained within the Standard Model or the CP-conserving MSSM, since it may have a 
weaker coupling to vector-boson pairs~\cite{CEPW}.

Extensions of the MSSM can be introduced, where the Higgs sector is
further enlarged and the Higgs masses are less constrained. We mention
here a single example, the so-called Next-to-Minimal Supersymmetric
Standard Model (NMSSM), whose Higgs sector includes not only two Higgs
doublets, but also an additional singlet~\cite{fayet}. The
phenomenology of the NMSSM Higgs sector was studied in \cite{EGHRZ}
(for an updated account of later developments, see also
\cite{CPHiggs}). Such an extension may slightly decrease the level of
fine-tuning required to reconcile the present stringent lower bounds
on supersymmetric particle and Higgs boson masses with the measured
value of the Fermi scale. With an additional singlet, the quartic
Higgs coupling becomes an independent parameter, as in the SM, thus
the NMSSM is less predictive than the MSSM: the MSSM upper limit on
the lightest Higgs boson mass is somewhat relaxed in the NMSSM,
becoming similar to the triviality bound of the SM. However, in the
NMSSM there is less need to push the mass parameters controlling
radiative corrections to extreme values, in order to satisfy the
present experimental bounds. 

\section{Prospects}

The Higgs boson may be not only the capstone of the Standard Model,
but also provide the opening to a whole new world beyond the Standard
Model. Much is known about its possible properties in both the
Standard Model and many supersymmetric scenarios. The stage is now set
for the LHC to prove all or most of these ideas wrong.



\section*{Acknowledgements}
This work was supported in part by the European Commission under the RTN 
program MRTN-CT-2004-503369.
%

\end{document}